\documentclass[aps,prl,twocolumn,superscriptaddress,showpacs]{revtex4}
\usepackage{graphicx}
\usepackage{rotating,dcolumn}
\begin{document}

\title{Hidden magnetic frustration by quantum relaxation in anisotropic Nd-langasite}

\author{V. Simonet}
\affiliation{Institut N\'{e}el, CNRS/UJF, B.P. 166, 38 042
Grenoble Cedex 9, France}
\author{R. Ballou} \affiliation{Institut N\'{e}el, CNRS/UJF, B.P. 166, 38 042 Grenoble Cedex 9, France}
\author{J. Robert}
\affiliation{Institut N\'{e}el, CNRS/UJF, B.P. 166, 38 042
Grenoble Cedex 9, France}
\author{B. Canals}
\affiliation{Institut N\'{e}el, CNRS/UJF, B.P. 166, 38 042
Grenoble Cedex 9, France}
\author{F. Hippert}
\affiliation{LMGP (CNRS), Grenoble Institute of Technology, 38016
Grenoble, France}
\author{P. Bordet}
\affiliation{Institut N\'{e}el, CNRS/UJF, B.P. 166, 38 042
Grenoble Cedex 9, France}
\author{P. Lejay}
\affiliation{Institut N\'{e}el, CNRS/UJF, B.P. 166, 38 042
Grenoble Cedex 9, France}
\author{P. Fouquet}
\affiliation{Institut Laue-Langevin, BP 156, 38042 Grenoble Cedex
9, France}
\author{J. Ollivier}
\affiliation{Institut Laue-Langevin, BP 156, 38042 Grenoble Cedex
9, France}
\author{D. Braithwaite}
\affiliation{CEA-Grenoble, INAC/SPSMS/IMAPEC, 17 rue des Martyrs,
38054 Grenoble cedex, France}

\date{\today}

\begin{abstract}
The static and dynamic magnetic properties of the
Nd$_3$Ga$_5$SiO$_{14}$ compound, which appears as the first
materialization of a rare-earth kagome-type lattice, were
re-examined, owing to contradictory results in the previous
studies. Neutron scattering, magnetization and specific heat
measurements were performed and analyzed, in particular by fully
taking account of the crystal electric field effects on the
Nd$^{3+}$ ions. One of the novel findings is that the peculiar
temperature independent spin dynamics observed below 10~K
expresses single-ion quantum processes. This would short-circuit
the frustration induced cooperative dynamics, which would emerge
only at very low temperature.
\end{abstract}


\maketitle

The kagome antiferromagnet is the prototype of geometrically
frustrated systems in 2 dimensions. It is predicted that its
ground state, in the classical case, remains disordered and is
highly degenerate \cite{KGM}. The investigation of such a
spin-liquid state is still impeded by the few available
experimental materializations. Various secondary perturbations
actually lead, in most known compounds, to long-range order (LRO)
or spin glass phase. This occurs, though, at a temperature well
below the onset of magnetic correlations and is often accompanied
by a dynamical behavior persisting below the transition
temperature. These dynamics are characterized by a slowing down of
the magnetic fluctuations which rounds off onto a relaxation
plateau \cite{mutka}. Although rather generic, the origin of this
peculiar relaxation is still debated.

In all the kagome compounds so far studied the magnetic entities
that form the frustrated lattice of corner-sharing triangles are
3d transition metal ions with weak magneto-crystalline anisotropy.
Inversely, in the 3 dimensional compounds of the pyrochlore
family, 4f rare-earth ions with strong magneto-crystalline
anisotropy occupy a frustrated lattice of corner-sharing
tetrahedra. This leads to a wider variety of magnetic behaviors.
Most striking, for instance, was the discovery of the spin ice
state in Ho$_2$Ti$_2$O$_7$, where the frustration arises from the
combined effects of multi-axial anisotropy and ferromagnetic
interactions \cite{spinice}. It was expected, in contrast, that
the multi-axial anisotropy would release the geometric frustration
of antiferromagnetic interactions in Tb$_2$Ti$_2$O$_7$, but this
compound exhibits no LRO and, instead, shows intriguing spin
dynamics well below the Curie-Weiss (CW) temperature
\cite{gardner,gingras}. It was recently argued that this compound
would form a quantum spin ice from dynamical frustration induced
through crystal field (CF) excitations and quantum many-body
effects \cite{molovian}.

The lately discovered Nd$_3$Ga$_5$SiO$_{14}$ (NGS) compound, which
belongs to the langasite family, provides with the unique
opportunity to study the role of a strong magneto-crystalline
anisotropy in a geometrically frustrated 2 dimensional lattice. In
this compound the rare-earth Nd$^{3+}$ ions fully occupy a lattice
showing the kagome corner-sharing triangles connectivity if only
first neighbor interactions are taken into consideration
\cite{bordet}.

\begin{figure}[t]
\includegraphics[bb=40 20 800 590,scale=0.32]{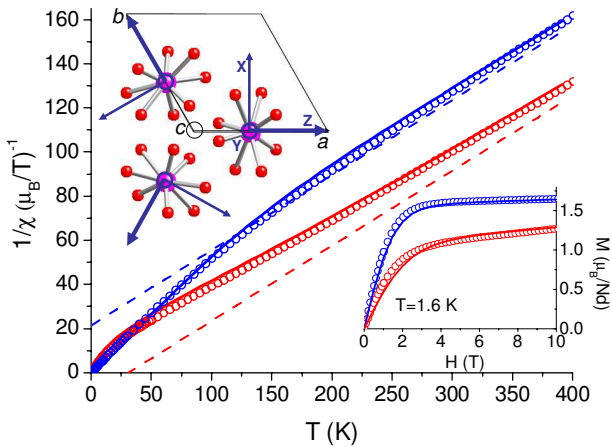}
\caption{(Color online) Full CF calculation (solid lines) of the
inverse susceptibility (main frame) and magnetic isotherms (inset)
with the magnetic field parallel (red) or perpendicular (blue) to
the kagome planes compared to single-crystal measurements
(circles) \cite{bordet}. The dashed lines are high temperature
expansion calculations. The three Nd$^{3+}$ are shown in their
local frame with their first coordination O$^{2-}$ shells.}
\label{CF}
\end{figure}

The magnetic properties of NGS powder samples and single crystal
were investigated in the 1.6$-$400~K temperature range
\cite{bordet}. Neutron diffraction and magnetization measurements
on the powders confirmed that NGS does not order in a N\'eel phase
nor shows any signature of a spin glass behaviour. Magnetization
measurements on the single-crystal (cf.~Fig.~\ref{CF}) revealed a
strong magneto-crystalline anisotropy associated with CF effects
on the ground multiplet J=9/2 of the Nd$^{3+}$ ions. A mean field
quantitative analysis of the magnetic susceptibility at high
temperature taking into account only the quadrupolar contribution
to the CF yielded a CW temperature $\theta=-52$~K, leading to
anticipate rather strong antiferromagnetic interactions between
the Nd moments. These were described as coplanar rotators in the
kagome planes. A change in the anisotropy occurs around 33~K,
below which the c-axis (perpendicular to the kagome planes)
becomes the macroscopic magnetization axis.

\begin{figure}[t]
\includegraphics[bb=0 160 814 572,scale=0.3]{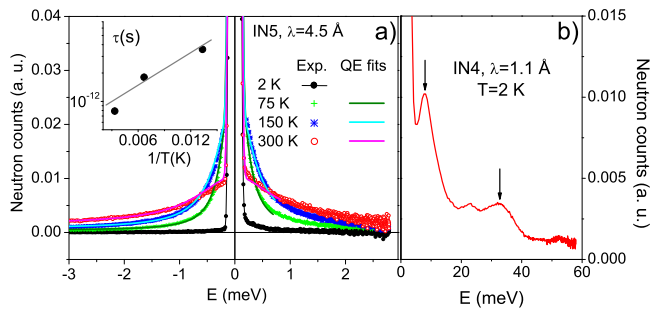}
\caption{(Color online) Time-of-flight spectra, $Q$-integrated in
the range [0.34, 2.55] \AA$^{-1}$. {\bf a)} Quasielastic magnetic
scattering fitted from 75 to 300 K by a Lorentz function
multiplied by the detailed balance factor. The fluctuation time
obtained from the inverse HWHM is shown in the inset together with
an Arrhenius fit ($\tau=\tau_0\exp(\Delta E/k_BT)$ with
$\Delta$E=133 K and $\tau_0$=6.7 10$^{-13}$ s). {\bf b)} Two
non-dispersive crystal-field level transitions pointed out by the
arrows. Note that the feature at $\approx$ 23 meV was attributed
to phonons from the $Q$-dependence.} \label{tof}
\end{figure}

Recently, a neutron scattering investigation of NGS at very low
temperature was reported \cite{zhou}. A weak ring of diffuse
magnetic scattering was observed at 46 mK, whose momentum-transfer
($Q$) dependence was found in agreement with calculated magnetic
scattering maps from spin-liquid models \cite{robert}. It was
inferred from additional neutron spin-echo (NSE) experiments, that
the associated magnetic state would be highly dynamical. The
authors also claimed to have evidenced a magnetic field induced
phase transition featured by a plateau of reduced magnetization,
approximately half the moment value of the free Nd$^{3+}$ ion, in
agreement with magnetization measurements (1.6~$\mu_B$ at 1.6 K)
\cite{bordet}. Another recent investigation by $\mu$SR and NMR
spectroscopy \cite{zorko} established that the magnetic
fluctuations slow down as the temperature is decreased to reach a
relaxation plateau below 10 K, interpreted as a signature of
magnetic frustration in this cooperative system. Some usual
signatures of frustration therefore appeared to have been observed
in NGS, magnetic spatial correlations and a fluctuating ground
state, but inconsistencies soon emerge from close inspection. In
reference \onlinecite{zhou}, no magnetic signal is reported at 46
mK outside an energy window corresponding to fluctuating times
larger than 5 10$^{-11}$ s but NSE experiments at the same
temperature appears to suggest a relaxation faster than
4~10$^{-12}$ s. In addition, this fast relaxation suggested by the
NSE experiment is incompatible with the characteristic fluctuation
time of the $\mu$SR-probed relaxation plateau, $\tau\approx~$2.5
10$^{-10}$ s. This motivated our experimental re-investigation of
the spin dynamics in NGS, presented below, as well as a full
characterization of the single-ion anisotropy through a complete
CF analysis.

Powders of Nd-langasite, synthesized as in ref.
\onlinecite{bordet}, were used for magnetization and specific heat
measurements and neutron scattering experiments. The spin dynamics
were investigated via time-of-flight (ToF) and NSE spectroscopy.
The ToF experiments were performed at ILL on the IN4 spectrometer
at a wavelength of 1.1~\AA~ and on the IN5 spectrometer with
wavelengths between 8.5 and 2.25~\AA. The ToF spectra reveal a
quasielastic (QE) signal, displayed in Fig.~\ref{tof}, whose
magnetic origin was inferred from its $Q$-dependence (slight
decrease following the Nd$^{3+}$ squared form factor). The half
width at half maximum (HWHM) of the $Q$-integrated spectra
decreases from 300~K to 75~K. An Arrhenius law can be forced to
account for the temperature dependence of the relaxation time
$\tau=\hbar/$HWHM. This yields an energy barrier of the order of
130~K for a thermally activated process. At 2~K, no QE signal can
be detected any more within the instrumental resolution energy
window.

\begin{figure}[t]
\includegraphics[bb=0 0 572 814, scale=0.35]{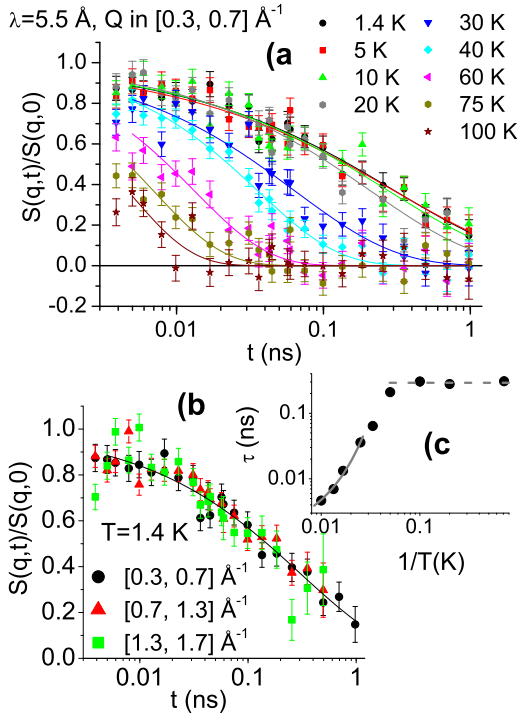}
\caption{(Color online) Neutron spin-echo normalized intermediate
functions fitted with stretched exponentials (solid lines) for
several temperatures ({\bf a}), and for several $Q$ ranges at 1.4
K ({\bf b}). The variation of the relaxation time over the inverse
temperature is displayed with an Arrhenius fit (solid line) at high
temperatures ($\Delta$E=135 K, $\tau_0$=1.2~10$^{-12}$~s)
and a dashed line underlying the relaxation plateau below
10 K ({\bf c}).} \label{NSE}
\end{figure}

In order to extend the time window for the observations of the
dynamics, a NSE experiment was performed on the IN11C spectrometer
at ILL using a wavelength of 5.5~\AA. XYZ polarization analysis
was performed to experimentally select the magnetic part of the
signal. The instrumental resolution was corrected with the purely
elastic magnetic signal of a reference compound. The accessible
time window of the NSE technique overlaps the ToF range for short
times and the $\mu$SR range for long times. The normalized
intermediate functions $S(Q,t)/S(Q,0)$, measured between 1.4 and
100~K, can be fitted by stretched exponential line-shapes
$\exp(-(t/\tau)^\beta)$ (cf.~Fig.~\ref{NSE}). A $\beta$ exponent
$\neq 1$ denotes a non-unique relaxation time. It increases
gradually from a constant value of 0.5 between 1.4 and 10~K to 0.8
at 40~K, indicating a change in the dynamics. The small temporal
range with non-zero signal at higher temperatures prevented from
obtaining a reliable value of $\beta$ which was therefore fixed
equal to 0.8 between 60 and 100 K. Above 40~K, the thermal
variation of the relaxation time agrees with a thermally activated
process ($\ln(\tau) \propto 1/T$). We get an energy barrier very
close to the one extracted from the ToF QE signal
(cf.~Fig.~\ref{NSE}) \cite{footnote1}. Below 10 K, the relaxation
time does no more vary with temperature, revealing a relaxation
plateau with a characteristic time $\tau\approx 3~10^{-10}$~s, in
good agreement with the $\mu$SR results \cite{zorko} (and at
variance with the NSE data of ref. \onlinecite{zhou}). An
important finding of our NSE experiment is that the magnetic
relaxation does not show any detectable $Q$-dependence in the
range [0.3 \AA$^{-1}$, 1.7 \AA$^{-1}$], pointing out the absence
of spatial correlation down to 1.4 K since the frequency width is
inversely proportional to the $Q$-dependent susceptibility in
correlated paramagnet \cite{lovesey}. The spatial correlations
observed at 46 mK \cite{zhou} thus would occur in a temperature
range much below the onset of the relaxation plateau.

\begin{figure}[t]
\includegraphics[bb=0 160 814 572,scale=0.3]{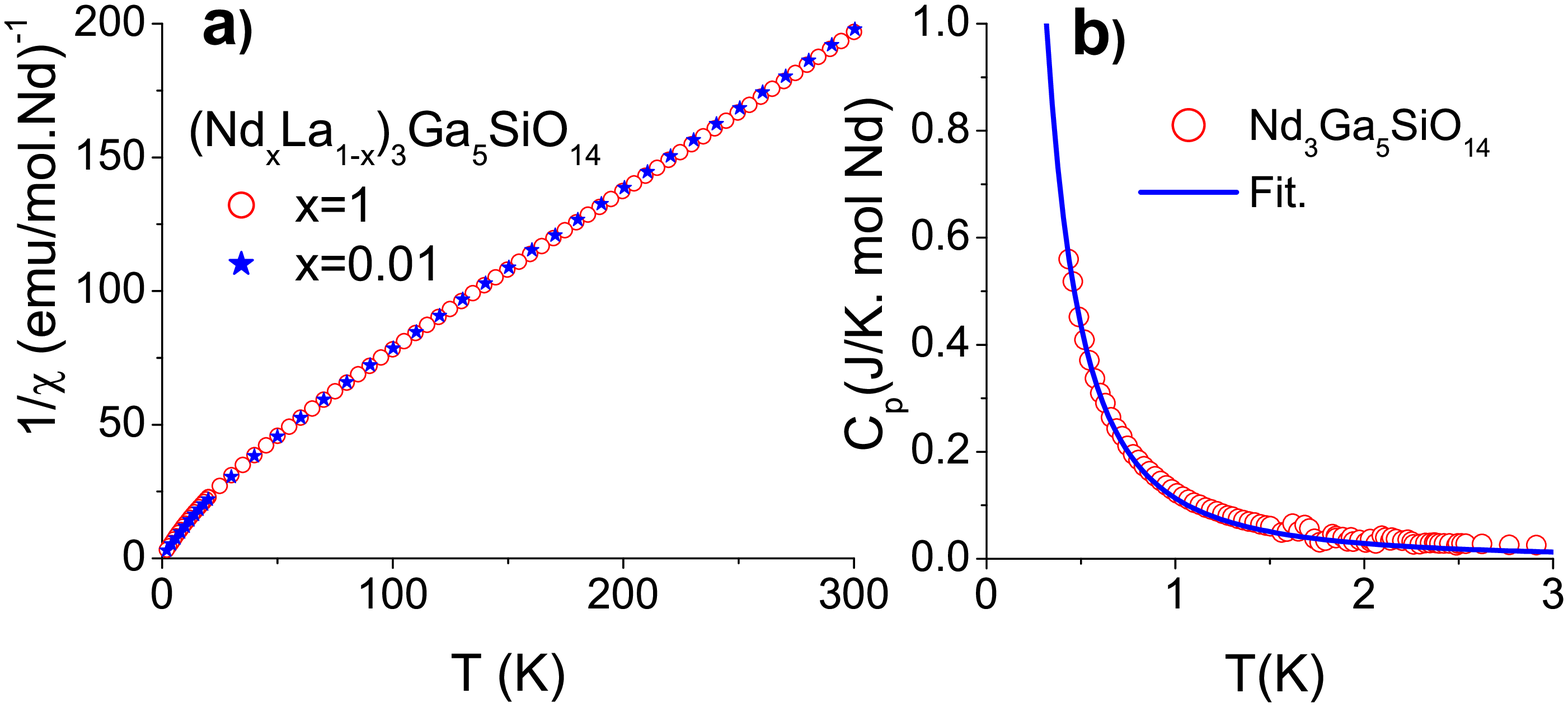}
\caption{(Color online) {\bf a)} Inverse susceptibility of NGS and
of a magnetically diluted sample. {\bf b)} NGS low temperature
specific heat fitted by $\rm R({\Delta\rm E\over \rm{k_BT}})^2
{\exp(-{\Delta\rm E\over \rm{k_BT}})\over (1+\exp(-{\Delta\rm
E\over \rm{k_BT}})^2}$ for a Schottky anomaly produced by a
two-level system with $\rm R$ the gas constant and $\Delta$E=235
mK.} \label{CvDil}
\end{figure}

Additional insights about the magnetism of NGS were gained from
the ToF scattering at higher energy transfer, specific heat
measurements and investigations of magnetically diluted samples.
Two CF level transitions can be identified, through their $Q$
dependence in the ToF spectra, around 95~K (also reported in
ref.~\onlinecite{zhou}) and 380~K (cf.~Fig.~\ref{tof}). They
however are very broad in energy, with a width at least one order
of magnitude larger than the instrumental resolution. Quantitative
estimations of CF effects suggest that this may be ascribed to the
distribution of the charge environments associated with the
Si$^{4+}$/Ga$^{3+}$ mixed site \cite{bordet}, which also is
responsible for the NMR line broadening reported in
ref.~\onlinecite{zorko}. The specific heat measurements were
performed down to 400 mK in a Quantum Design PPMS. The magnetic
contribution was determined after subtraction of the specific heat
measured on the isostructural non-magnetic La$_3$Ga$_5$SiO$_{14}$
compound. It exhibits a broad bump ranging from $\approx$100~K to
above 300~K. A strong rise is noticeable below 2~K  as the
temperature is decreased down to the lowest values
(cf.~Fig.~\ref{CvDil}). We shall discuss this below. Also
meaningful were the magnetization measurements on the magnetically
diluted (Nd$_{0.01}$La$_{0.99}$)$_3$Ga$_5$SiO$_{14}$ compound,
which were performed in order to isolate the influence of the
magneto-crystalline anisotropy. We found out, to our surprise,
that its magnetic susceptibility does experimentally not differ
from that of NGS (cf.~Fig.~\ref{CvDil}), demonstrating that the
exchange interaction between the Nd$^{3+}$ moments in NGS is
certainly much smaller than initially estimated \cite{bordet}, at
least not greater than a few K.

The above results were confronted to calculations of the CF
effects. In NGS, there are 3 distinct Nd$^{3+}$ ions per unit
cell, each lying on a different 2-fold axis, inter-related through
the  3-fold $\vec c$ axis of the crystal \cite{bordet}, thus
giving rise to multi-axial low symmetry CF potentials. The
associated Hamiltonian for each Nd$^{3+}$ ion in the ground
multiplet $J$=9/2 can be expressed as
$H^{CF}=\sum_{p=1}^3{\sum_{q=-p}^p{A_{2p}^{2q}O_{2p}^{2q}}}$ with
$A_{2p}^{-2q} = (A_{2p}^{2q})^{\star}$, in terms of $O_{n}^{m}$
Racah operators and with respect to a local frame where the 2-fold
axis is the quantization axis for the angular momentum \cite{CEF}.
Ignoring exchange interactions this provides with an energy
spectrum which can be compared with the transitions detected in
the ToF spectra and can be used to calculate the specific heat.
The magnetization and susceptibility are obtained from the
eigenfunctions and must be averaged over the 3 Nd$^{3+}$ ions.
This forced us to work in a common global frame with the $\vec c$
axis as the quantization axis and to perform appropriate rotations
of the Racah operators. It was illusory to try to determine the 9
independent $A_{2p}^{2q}$ coefficients at once. Thus, in a first
step the $A_{2p}^{0}/A_{2p}^{2q}$ ratios were calculated in the
point charge model from the first coordination shell of eight
O$^{2-}$ (cf.~Fig.~\ref{CF}). Qualitatively good results were
already obtained, reproducing in particular the crossing of the
susceptibility for applied fields parallel and perpendicular to
$\vec c$. This crossing is due to the 4$^{\rm th}$ order terms of
$H^{CF}$. The $A_{2p}^{2q}$ coefficients were, in a second step,
slightly varied in order to better account for all the
experimental constrains (details of the calculations will be given
elsewhere). With the best set of $A_{2p}^{2q}$ parameters, the
energy spectrum consists of 5 degenerate doublets, the first two
being separated from the ground state doublet by 90~K and 324~K
respectively, in rather good agreement with the ToF spectra
considering the width of the measured excitations. The ground
state in the global frame with the $\vec c$ quantization axis is a
combination of $|\pm 7/2>$, $|\pm 5/2>$ and $|\pm 3/2>$ states
with a maximum weight for the $|\pm 5/2>$ state. The good
agreement between the calculated and measured magnetic isotherms
and susceptibility is shown in Fig. \ref{CF}. A first observation
is that the low temperature reduced moment must be attributed to a
CF effect, the so-called field-induced magnetic phase
\cite{zhou,zorko} being hence a polarized paramagnet, at least at
1.6 K, as in Tb$_2$Ti$_2$O$_7$ \cite{rule}. A second observation
is that the susceptibility reaches the high temperature behaviour
above 2000 K. This regime occurs thus at a temperature much higher
than the one associated to the highest crystal field level energy
(665 K). This is a consequence of the large 6$^{\rm th}$ order
terms, shifting the inverse susceptibilities upward, which
explains why the previous estimation of the exchange interactions
went erroneous despite the almost linear variations of the inverse
susceptibility \cite{bordet}.

The CF analysis confirms the smallness of the exchange interaction
in consistency with the absence of spatial correlations down to
1.4~K. This suggests a single-ion origin for the measured magnetic
relaxation both in the varying and independent temperature
regimes, as probed by the neutron spectroscopy. The first excited
doublet of the calculated CF spectrum lies at an energy in
qualitative agreement with the barrier extracted from the ToF and
NSE measurements at temperatures above 40 K. The thermally
activated dynamics in this temperature range are most probably
mediated by the spin-phonon interaction \cite{villain2}. On
decreasing the temperature, this progressively slows down, but
below 10~K is short-circuited by another process with a
temperature independent relaxation time. This strongly suggests
quantum relaxation phenomena \cite{mol,LiHo} and calls to mind the
dynamical cross-over observed in spin ice compounds
\cite{ehlers2003}. The origin of this relaxation is still unclear
but could be related to the only feature not accounted for by the
CF calculation: the rise of the specific heat at very low
temperature. A splitting of the ground state doublet was suggested
in ref. \onlinecite{zhou} from extrapolation of Tof data to
zero-field under magnetic field. Assuming this suggestion, we
fitted the specific heat by a $T^{-2}$ law in agreement with the
tail of a Schottky anomaly associated with a two-level system (cf.
Fig. \ref{CvDil}). This gives an energy gap of $\approx$235~mK, in
good agreement with the reported zero-field extrapolated splitting
\cite{zhou}. Although the cause of this splitting is unidentified,
we know that it should necessarily break the time reversal
symmetry, since the Nd$^{3+}$ are Kramers ions. An interaction
involving odd powers of the ladder operators of the Nd$^{3+}$
angular momentum would do, by providing with non-diagonal matrix
elements mixing the two orthogonal states of the doublet. The
consequences of this low temperature dynamics on the cooperative
magnetic phase, expected when the exchange interactions and the
geometric frustration become significant, can now be addressed. A
cooperative dynamical behaviour presenting longer fluctuation
times than the fluctuations created by the proposed single-ion
quantum process would indeed be hindered. This, thus, queries the
nature of the phase associated to the ring-like $Q$-distribution
at 46~mK \cite{zhou}.

In conclusion, the experimental re-investigations of the
kagome-like rare-earth NGS compound combined with CF analysis have
allowed featuring its magnetism on firm grounds, solving previous
inconsistencies. The interpretation of the relaxation plateau of
the moment dynamics  in terms of cooperative processes is ruled
out. Its origin must rather be sought from single-ion quantum
processes. The effects of the geometrical frustration then would
be hidden emerging only at very low temperature.

\acknowledgments This work was financially supported by the ANR
06-BLAN-01871.


\begin{thebibliography}
\bibitem{footnote1} The factor 2 in the $\tau_0$ values obtained from
the Tof and NSE data may be due to the distribution of relaxation
times. This is revealed from the fits of the NSE data by stretched
exponentials whose Fourier Transforms are not rigorously the
Lorentz functions used in the Tof analysis.
\end{thebibliography}
\end{document}